 \documentstyle[multicol,aps,prl]{revtex}
 \renewcommand{\narrowtext}{\begin{multicols}{2} \global\columnwidth20.5pc}
 \renewcommand{\widetext}{\end{multicols} \global\columnwidth42.5pc}
\input{epsf.tex}

\def\top#1{\vskip #1\begin{picture}(290,80)(80,500)\thinlines \put(
65,500){\line( 1, 0){255}}\put(320,500){\line( 0, 1){
5}}\end{picture}}
\def\bottom#1{\vskip #1\begin{picture}(290,80)(80,500)\thinlines \put(
330,500){\line( 1, 0){255}}\put(330,500){\line( 0, -1){
5}}\end{picture}}

\begin{document}
\title{Wavefunctions and counting formulas for quasiholes\\ of
clustered quantum Hall states on a sphere}
\author{N. Read}
\address{Department of Physics, Yale University, P.O.\ Box
208120, New Haven, CT 06520-8120}

\date{May 5, 2006}
\maketitle
\newcommand{\be}{\begin{equation}}
\newcommand{\ee}{\end{equation}}
\newcommand{\bea}{\begin{eqnarray}}
\newcommand{\eea}{\end{eqnarray}}
\newcommand{\non}{\nonumber}
\begin{abstract}
The quasiholes of the Read-Rezayi clustered quantum Hall states
are considered, for any number of particles and quasiholes on a
sphere, and for any degree $k$ of clustering. A set of trial
wavefunctions, that are zero-energy eigenstates of a $k+1$-body
interaction, and so are symmetric polynomials that vanish when any
$k+1$ particle coordinates are equal, is obtained explicitly and
proved to be both complete and linearly independent. Formulas for
the number of states are obtained, without the use of (but in
agreement with) conformal field theory, and extended to give the
number of states for each angular momentum. An interesting
recursive structure emerges in the states that relates those for
$k$ to those for $k-1$. It is pointed out that the same numbers of
zero-energy states can be proved to occur in certain
one-dimensional models that have recently been obtained as limits
of the two-dimensional $k+1$-body interaction Hamiltonians, using
results from the combinatorial literature.
\end{abstract}

\narrowtext %

Non-Abelian quantum Hall states have been a subject of great
interest for some time \cite{mr}. Moore and the author showed that
non-Abelian statistics, in which adiabatic exchange of
well-separated quasiparticles produces a matrix operation on a
space of degenerate states, was a possibility in a condensed
matter system, and gave an example, now often termed the
Moore-Read (MR) state. They related the deep structure underlying
the statistics and other aspects of the quasiparticles of these
phases to conformal field theory, and to what are now called
modular tensor categories. Many trial wavefunctions of quantum
Hall states can be related to conformal blocks (i.e.\ chiral parts
of correlation functions) in the corresponding conformal field
theory. A further sequence of examples was constructed by Rezayi
and the author (RR) \cite{rr99}. These are parametrized by two
integers, $k\geq 1$ and $M\geq0$; the filling factor for each is
$\nu=k/(Mk+2)$, and they describe bosons for $M$ even, fermions
for $M$ odd (the case $k=1$ is the Laughlin state, while $k=2$ is
the MR state). For $M=0$ or $1$, trial ground state and quasihole
states can be obtained as zero-energy eigenstates of a Hamiltonian
that consists of a purely $k+1$-body interaction, acting within
the lowest Landau level. (For $M>0$, the trial wavefunctions are
those for $M=0$ multiplied by the factor $\prod_{i<j}(z_i-z_j)^M$,
where $z_i=x_i+iy_i$ is the complex coordinate representing
position $(x_i,y_i)$ of particle $i$ in two dimensions. These
trial wavefunctions are zero-energy eigenstates for a Hamiltonian
that contains two-body as well as $k+1$ body interactions
\cite{rr99}.) For $M=0$, the wavefunctions are closely related to
conformal blocks of SU(2) level $k$ current algebra conformal
field theories \cite{rr99}. While the explicit trial ground state
wavefunctions were obtained, for $k\geq3$ the trial wavefunctions
for quasiholes were not, except for some of those for small
numbers of quasiholes \cite{rr99,cgt}. The number of degenerate
zero-energy states for fixed positions of the quasiholes is basic
to the theory of non-Abelian statistics, and for the MR and RR
states these degeneracies have been studied numerically and using
conformal field theory methods
\cite{nayak,rr96,rr99,gr,ardonne,rj}.

The MR state is now believed theoretically to describe the phase
observed at filling factor $\nu=5/2$ in electron systems
\cite{morfrh}, while there is evidence that the RR $k=3$, $M=1$
state occurs at $\nu=12/5$ \cite{rr99}. For bosons, there is
numerical evidence that the sequence of RR states is relevant to
rotating cold, trapped, bosonic atoms \cite{cwg,rrc}. Interest in
non-Abelian statistics is presently being fueled by the
possibility of topological quantum computation \cite{freedman}.

In this paper, we return to the question of finding and counting
the zero-energy states of the $k+1$-body interaction for bosons,
for general $k$. We work in the lowest Landau level on the sphere,
with $N_\phi$ quanta of magnetic flux piercing it \cite{hald83}.
The approach is a generalization of one used to prove linear
independence of a proposed set of functions for the $k=2$ case in
Ref.\ \cite{rr96}, and is very close to work \cite{sf} on the same
counting problem for the limit $N_\phi\to\infty$, which has been
reviewed, extended, and applied in Ref.\ \cite{aks}. (The authors
of the first paper in Ref.\ \cite{sf} discovered the relation of
the symmetric functions that vanish when $k+1$ coordinates
coincide to conformal blocks of SU(2) current algebra at level
$k$, which was discussed independently in Ref.\ \cite{rr99}.
{}From the quantum Hall point of view, the results on the
$N_\phi\to\infty$ case describe edge excitations.) Our main
results are: (i) a complete and linearly-independent set of
explicit wavefunctions for zero-energy states of the $k+1$-body
interaction for any number of particles $N$ and any number of flux
quanta $N_\phi$, or alternatively, any number $n$ of quasiholes,
as defined by%
\be%
N_\phi=\frac{2}{k}N+M(N-1)+\frac{n}{k}-2;\ee%
(ii) explicit generating functions that encode the number of
zero-energy states for any $N$, $n$, decomposed according to
orbital angular momentum $L_z$ (and hence $L$ also); (iii) an
interesting recursive structure emerges, that relates the
quasihole problem for $k$ to the similar problem for $k-1$ at a
reduced number of particles and flux quanta, and might be useful
in making further progress in understanding the quasihole states.
The arguments are short and self-contained, and make no use of
conformal field theory. In an appendix, we also consider quasihole
wavefunctions of a natural form that generalizes those in Ref.\
\cite{cgt}, but for these our results are less conclusive.

We want to find the complete set of zero-energy states for the
$k+1$-body Hamiltonians in Ref.\ \cite{rr99}. The cases with $M>0$
can be found by multiplying those for $M=0$ by the
Laughlin-Jastrow factor $\prod_{i<j}(z_i-z_j)$ to the power $M$,
so we can focus on the $M=0$ case (hence, on bosons). Then the
problem reduces to the following: find all the symmetric
polynomials in $N$ complex variables $z_i$ that vanish when any
$k+1$ of the $z_i$ are equal. For the present, working on the
sphere means that the polynomial is of degree at most $N_\phi$ in
each $z_i$ \cite{hald83}. We will denote the vector space of
symmetric polynomials with these properties by $V_{N,N_\phi,k}$
and a typical member of this space by
$\widetilde{\Psi}_{N,N_\phi,k}$ (the polynomials
$\widetilde{\Psi}$ can be related to wavefunctions $\Psi$ on the
sphere or plane as explained in Ref.\ \cite{rr96}).

Consider a function $\widetilde{\Psi}_{N,N_\phi,k}$. While it
vanishes whenever $k+1$ particles coincide, it may not vanish when
only $k$ coincide. Let $z_1$, $z_2$, \ldots, $z_k$ tend to the
same value, which we will denote by $Z_1$. If it vanishes in this
limit, then (since it is totally symmetric) it vanishes when any
$k$ coordinates coincide, and belongs to $V_{N,N_\phi,k-1}$. If it
does not vanish, then its value (corresponding to the ``residue''
in Ref.\ \cite{rr96}) is a function of $Z_1$ and of $z_{k+1}$,
\ldots, $z_N$, and is symmetric in the latter. In this residue we
may let another set of $k$ coordinates, say $z_{k+1}$, \ldots,
$z_{2k}$ coincide at $Z_2$. If the function vanishes in this
second limit, then the residue consists of a function of $Z_1$
times a member of $V_{N-k,N_\phi,k-1}$. If it does not vanish in
the second limit, we obtain a second residue, to which the
procedure can be applied again.

In this way the space of functions is divided into a nested
sequence of subspaces within subspaces (called a ``filtration'' in
Ref.\ \cite{aks}). Each such subspace contains all the
wavefunctions that have a non-vanishing residue, which is a
function of $Z_1$, $Z_2$, $Z_3$, \ldots, (each of which are the
locations of $(N-F)/k$ ``clusters'' of $k$ particles), and of the
remaining $F$ coordinates $z_{N-F+1}$, \ldots, $z_N$ of the $F$
``unclustered'' particles; these residues vanish if any $k$ of the
latter set coincide. Each subspace also contains functions with
vanishing residues, which are also members of a subspace with
fewer clusters. Thus, the $(N-F)/k$th subspace [$(N-F)/k=0$, $1$,
\ldots ] can be defined as the space of functions that vanish
identically if $(N-F)/k+1$ clusters of $k$ particles each coincide
at $Z_1$, \ldots, $Z_{(N-F)/k+1}$, for all values of these
coordinates and of the remaining $z$'s. We will show that the
non-vanishing residues of functions in this subspace must take
the following form:%
\widetext
\top{-2.8cm}
\be%
\widetilde{\Psi}_{N,N_\phi,k}\to \prod_{p<p'}(Z_p-Z_{p'})^{2k}
\prod_{p'',l}(Z_{p''}-w_l)\prod_{p''',m} (Z_{p'''}-z_{N-F+m})^2
\widetilde{\Psi}_{F,N_\phi-2(N-F)/k,k-1}.\label{res}\ee%
\bottom{-2.7cm} \narrowtext \noindent %
The notation and the justification for each of the factors is as
follows. The first product is over pairs of cluster coordinates
$Z_p$, labeled by $p$, $p'$, \ldots, $=1$, $2$, \ldots, $(N-F)/k$
(the number of clusters). The function vanishes when any $k+1$
particles coincide, so it must vanish when any two clusters
coincide, and since it is symmetric under exchanges of clusters,
it must contain a factor $(Z_{p}-Z_{p'})^2$ for each pair of
clusters. In fact, because the original wavefunctions are
symmetric functions for particles on the sphere, it must vanish as
the $2k$th power; the proof of this is given in an appendix to
this paper. The last factor
$\widetilde{\Psi}_{F,N_\phi-2(N-F)/k,k-1}$ is a function only of
the $F$ ``unclustered'' coordinates, so is independent of the
cluster coordinates. By assumption, it vanishes if any $k$ of the
remaining $F$ coordinates coincide, so can be any function in
$V_{F,N_\phi-2(N-F)/k,k-1}$. The residue must also vanish if any
of the unclustered particles approaches a cluster, so it must
contain $Z_{p}-z_{N-F+m}$ for the $p$th cluster and the $m$th
unclustered particle ($m=1$, $2$, \ldots, $F$). In fact, it must
vanish quadratically, which again is proved in the Appendix. This
form implies that the maximum degree in the unclustered
coordinates in the remaining factor is $N_\phi-(N-F)/k$, as
recorded in the second subscript on
$\widetilde{\Psi}_{F,N_\phi-2(N-F)/k,k-1}$. Finally, the residue
can contain a further factor of a symmetric polynomial in the
cluster coordinates. As the degree in any of the cluster
coordinates is at most $kN_\phi$, it works out that the maximum
degree in each cluster coordinate in this symmetric polynomial
factor is $n$. Each quasihole effectively contains $1/k$ of a flux
quantum or vortex \cite{rr99}. In general, $n$ need not be
divisible by $k$ (in Ref.\ \cite{rr99}, only the special cases $N$
and $n$ divisible by $k$ were considered). For $N_\phi$ to be an
integer, we require that $2N+n$ be divisible by $k$ (hence for $k$
even, $n$ must be even), and ${\rm dim}\, V_{N,N_\phi,k}$ (the
dimension of $V_{N,N_\phi,k}$) is nonzero only when $n\geq0$. The
symmetric polynomials in the cluster coordinates of this degree
can be expressed using the product $\prod_{p'',l}(Z_{p''}-w_l)$ as
a generating function (by expanding it in symmetric polynomials in
the variables $w_l$). These are then ``coherent states'', in which
the coordinates $w_l$ can be interpreted as the positions of the
quasiholes. The function vanishes when any $k$ particles approach
a $w_l$, as it should \cite{arrs}, and this can be taken as a
definition of the quasihole coordinates for these quasiholes,
which have charge $-1/2$ (in units of particle number).

Let us suppose that the problem has already been solved for values
$k'=k-1$. Then we have a set of linearly independent functions
$\widetilde{\Psi}_{F,N_\phi-2(N-F)/k,k-1}$ that span
$V_{F,N_\phi-2(N-F)/k,k-1}$. Then by expanding in the quasihole
coordinates $w_l$, we have a set of non-vanishing residues that
are functions of the cluster coordinates and the unclustered
particle coordinates, and clearly are linearly independent. This
implies that if for each of these residues, there is a
wavefunction from which it arose by taking limits, then these
wavefunctions (which are in $V_{N,N_\phi,k}$) are linearly
independent. The existence of a wavefunction that reduces to each
non-vanishing residue in the limit will be shown in a moment.

Completeness of the set of functions can also be proved by
induction. As the form of the non-vanishing residues has been
fixed, we must show that there is a one-one mapping from the
wavefunctions to the non-vanishing residues. This is done by
induction on the number of clusters. For the residues with no
clusters ($F=N$), the residue reduces to
$\widetilde{\Psi}_{N,N_\phi,k-1}$, for which a complete set is
assumed to have already been found. Then for the residues with one
cluster ($F=N-k$), if there are two wavefunctions with the same
residue, then the difference of these functions is a wavefunction
with no clusters (since it vanishes when $k$ particles coincide),
and these are also in the space $V_{N,N_\phi,k}$. This argument
can be repeated, and so by induction we find that there is (up to
addition of a wavefunction for fewer clusters) at most one
wavefunction for each non-zero residue, that is (assuming
existence) the dimension of the space of non-vanishing residues
and that of the space of wavefunctions are the same for each $N$,
$N_\phi$, and $k$.

To prove the existence of a zero-energy wavefunction that tends to
each non-vanishing residue in the limit, we construct them. We use
the following labeling for the particles: The particle number $N$
is partitioned into $m_\alpha\geq0$ clusters of size $\alpha=1$,
\ldots, $k$, that is $N=\sum_\alpha \alpha m_\alpha$ (there will
be a set of wavefunctions for each such partition). One particle
is assigned to each box in a Ferrers-Young diagram in which there
are at most $k$ columns, and which consists of $m_\alpha$ rows of
length $\alpha$ for each $\alpha$, arranged as usual with the row
lengths weakly decreasing as one goes down the diagram. Then
(following Ref.\ \cite{aks}) the particle coordinates will be
written $z^{(\alpha)}_{ij}$, where $i$, $j$ label rows and
columns, respectively (as for a matrix, $i$ increases down the
columns, and $j$ increases to the right along the rows) of the
rectangular block consisting of all rows of length $\alpha=1$,
\ldots, $k$. Thus for each $\alpha$, $j=1$, \ldots, $\alpha$, and
$i=1$, \ldots, $m_\alpha$. The rows of the diagram are ordered
with $(\alpha,i)$ above $(\beta,i')$, written
$(\alpha,i)>(\beta,i')$, if either $\alpha>\beta$, or
$\alpha=\beta$ and $i<i'$  (note the direction of the last
inequality). The wavefunctions will be written down in a form that
deals with all levels of the recursion at once. It will turn out
that $(N-F)/k=m_k$, while the remaining $m_\alpha$ represent,
similarly, the numbers of clusters of sizes $\alpha$ smaller than
$k$. Similarly, the wavefunctions involve not only quasihole
coordinates $w_l^{(k)}$ ($l=1$, \ldots, $n_k=n$), which can be
identified with the $w_l$ in the residues, but also further
coordinates $w_l^{(\alpha)}$ for all $\alpha=1$, \ldots, $k-1$,
which play a corresponding role in the parametrization of the
wavefunctions for smaller $k$ that appear in the recursion. Here
for each $\alpha$ the range is $l=1$, \ldots, $n_\alpha$, with
$n_1\leq n_2\leq\cdots \leq n_k$.

The wavefunctions are \widetext \top{-2.8cm}
\be%
\widetilde{\Psi}_{N,N_\phi,k}(z_1,\ldots,z_N;\{w^{(1)}_l\},\ldots,
\{w^{(k)}_l\})={\cal S}_z
\left\{\prod_{(\alpha,i)>(\beta,i')}\prod_{j=1,\ldots,\beta}
(z^{(\alpha)}_{i,j}-z^{(\beta)}_{i',j})
(z^{(\alpha)}_{i,j+1}-z^{(\beta)}_{i',j})\cdot\prod_\alpha\prod_{i,j}
{\prod_l}^{(\alpha,j)} (z^{(\alpha)}_{ij}-w^{(\alpha)}_l)
\right\}.\label{wfn}\ee%
In this expression, ${\cal S}_z$ is the symmetrizer over all the
$N$ particles, and $z^{(\alpha)}_{i,\alpha+1}=z^{(\alpha)}_{i,1}$.
In each product $\prod^{(\alpha,j)}_l$, $l$ runs over%
\be%
l=n_{j-1}+(1-\delta_{j,1})\sum_{\beta:j-1\leq\beta<\alpha}
m_\beta+1,\ldots,n_j
+\sum_{\beta:j-1<\beta<\alpha}m_\beta\ee%
\bottom{-2.7cm}
\narrowtext \noindent %
(where $n_0=0$). To ensure that each function is of the same
degree
$N_\phi=(2N+n)/k-2$ in all coordinates $z_i$, we require%
\be%
2\sum_{\beta:\beta\geq\alpha}
m_\beta+n_\alpha-n_{\alpha-1}=(2N+n)/k\ee%
for all $\alpha=1$, \ldots, $k$. Also, for each $\alpha$, the
wavefunctions are of the same degree $m_\alpha$ in
$w^{(\alpha)}_l$ for all $l=1$, \ldots, $n_\alpha$.

The structure of the $z-z'$ factors inside the symmetrizer is
based on that in Ref.\ \cite{aks}, and for the factors that
connect two rows $(\alpha,i)$, $(\beta,i')$ with $\alpha=\beta$ is
the same as in the ground states (i.e.\ the case $m_k=N/k$,
$m_\alpha=0$ for $\alpha<k$) in Ref.\ \cite{rr99}. These
particle-particle factors are of two types: one type connects two
particles (and vanishes when their coordinates coincide) whenever
they are in the same column in the diagram. The other type
connects two particles in different rows only if the one in the
higher row is one column to the right of the lower one; when the
rows are of the same length, this is interpreted cyclically along
the rows. The following arguments are a version of the arguments
for existence of a wavefunction mapping to each residue in Ref.\
\cite{aks}, and generalize an argument for the ground state
wavefunctions in Ref.\ \cite{rr99}.

The wavefunctions in eq.\ (\ref{wfn}) vanish when any $k+1$
coordinates are equal, because in each term inside the
symmetrizer, at least two must be in the same column of the
diagram (i.e.\ have the same value of $j$). They also vanish if
$k$ particles coincide at $Z_1$ say, unless the particles lie in
the same row (which must be of length $k$). First, these particles
must be in distinct columns for the wavefunction to have a
possibility of not vanishing, and hence one must be in column $k$,
and so in the first $m_k$ rows. Then in order for the function not
to vanish as the particles come together, the one in column $k-1$
must be in the same or a higher row, and so on, until in the first
column the particle will be in the same or a higher row as that in
column $k$. If the row is higher, the wavefunction will vanish
because of the cyclic connection of the first and last columns.
Hence all $k$ must lie in the same row. Repeating this argument as
another $k$ particles tend to $Z_2$, and so on, one finds that in
the clustered limit, when there are $m_k$ clusters with $k$
coinciding coordinates in each, the particles in each cluster came
from the same row in each term that contributes a non-vanishing
term to the residue. The residue vanishes if any $k$ of the
remaining coordinates coincide, and vanishes as $(Z_i-Z_j)^{2k}$
if any two clusters of $k$ coincide. The wavefunctions also vanish
if $k$ particles coincide with a $w^{(k)}_l$. Again, the $k$
particles must be in the same row, and then exactly one of them
appears in a factor $(z-w^{(k)}_l)$, so the wavefunction vanishes,
and the residue does likewise. Finally, the residue vanishes
quadratically as $(Z_i-z')^2$ for any of the remaining unclustered
coordinates $z'$, as each of them is connected to exactly two
particles in each row of length $k$. This establishes that the
residue is of the form in eq.\ (\ref{res}), with
$\widetilde{\Psi}_{F,N_\phi',k'}$ being a function in the
remaining $z$'s and $w$'s of the same form as in eq.\ (\ref{wfn}),
with parameters as indicated. Then arguing by induction on $k$
establishes that there is a wavefunction for each non-vanishing
residue, as claimed.

Taking all the arguments together, we can conclude that the
wavefunctions in eq.\ (\ref{wfn}), though written as overcomplete
sets of ``coherent states'' using auxiliary coordinates
$w^{(\alpha)}_l$, span a complete set of linearly-independent
zero-energy wavefunctions. Because the residues are invariant
under permutations of the $w^{(\alpha)}$'s among themselves (for
fixed $\alpha$), no states are lost if the wavefunctions are
symmetrized over the $w^{(\alpha)}$'s also, by applying
$\prod_\alpha {\cal S}_{w^{(\alpha)}}$. Then the functions can be
expanded as linear combinations of polynomials in the $z$'s times
products over $\alpha$ of symmetric polynomials in the
$w^{(\alpha)}_l$'s for each $\alpha$, and the polynomials in $z$
for each distinct product of symmetric polynomials in the $w$'s
form a complete and linearly independent set of wavefunctions.

For $k=1$, the wavefunctions are the Laughlin quasihole
wavefunctions. For $k=2$ they are the same complete set of
linearly-independent functions found in Ref.\ \cite{rr96} (and for
$N$ even and $n=2$ quasiholes, in Ref.\ \cite{mr}). In the latter,
the unclustered (unpaired) particles were viewed as fermions
occupying zero modes, but we now see that they can also be viewed
as Laughlin quasihole wavefunctions for $\nu=1/2$ involving $F$
particles and $n_1$ quasiholes. For general $k$, they agree with
Ref.\ \cite{rr99} for $n=k$, $N$ divisible by $k$.

We can now find the ``number of states'' $\#_{N,N_\phi,k}={\rm
dim}\,V_{N,N_\phi,k}$, by counting the non-vanishing residues (or
by counting the wavefunctions directly). In these first results,
we use only the recursive structure. We expand the symmetric
polynomials in the residues in eq.\ (\ref{res}) in the $w_l$'s, so
as to obtain linearly-independent residues; the number of
linearly-independent such functions is a binomial coefficient
\cite{rr96}, and this
then yields%
\be%
\#_{N,N_\phi,k}=\sum_{F:F\equiv N ({\rm
mod}\, k)}{\frac{N-F}{k}+n\choose n}\#_{F,N'_\phi,k'},\label{rec}\ee%
where $N'_\phi=N_\phi-2(N-F)/k=(2F+n)/k-2$, and $k'=k-1$. This has
the general structure found for $k=2$ in Ref.\ \cite{rr96}. For
general $k$, the structure was conjectured in Ref.\ \cite{rr99},
and results were obtained in \cite{gr,ardonne} (although the form
of the ``spatial degeneracy'' factors, i.e.\ the binomial
coefficients in this formula, seems never to have been derived
generally). In the literature, this form is
usually written with the notation \cite{gr,ardonne}%
\be%
\left\{\begin{array}{c}n\\F\end{array}\right\}_k=\#_{F,N'_\phi,k'},\ee%
when $n+2F$ is divisible by $k$. Remarkably, we have found that
these coefficients, which represent the ``internal'' degeneracy of
the quasiholes, are themselves related to a problem of counting
the full degeneracy of zero energy quasihole states, but at a
different value of $k$. This now allows an inductive solution for
these coefficients, recursively in $k$.

First, we can write a recursion relation for
the coefficients:%
\be%
\left\{\begin{array}{c}n\\F\end{array}\right\}_k=
\sum_{F':F'\equiv F({\rm mod}\,k')}{\frac{F-F'}{k'}+n'\choose n'}
\left\{\begin{array}{c}n'\\F'\end{array}\right\}_{k'},\ee%
where again $k'=k-1$, and%
\be%
n'= (k'n-2F)/k.\ee

Next we mention that for $k=1$, the residues above are just the
wavefunctions, and the function from the problem for $k'=0$ is
simply unity. Those wavefunctions are the Laughlin quasihole
functions. This implies that
$\left\{\begin{array}{c}n\\F\end{array}\right\}_1=\delta_{F,0}$.
In this case, the statistics is Abelian, and the only degeneracy
is due to the different possible locations of the quasiholes.
The recursion relation then yields for $k=2$,%
\be%
\left\{\begin{array}{c}n\\F\end{array}\right\}_2={n/2 \choose
F},\ee%
as found in Ref.\ \cite{rr96}. When these numbers, or their sum
over $F$, are larger than 1, they describe the degeneracy of
states for quasiholes even when their positions are fixed, a
characteristic of a non-Abelian state.

When these values are substituted into the recursion relation for
$k=3$, the result can be immediately rewritten%
\be%
\left\{\begin{array}{c}n\\F\end{array}\right\}_3=
\sum_{a,b:a+2b=F}{\frac{n-F}{3}\choose
a}{\frac{2n-(2a+b)}{3}\choose b},\ee%
which was found previously by Ardonne \cite{ardonne}, by a much
longer method using conformal field theory and results from Ref.\
\cite{schout}.

The solution for arbitrary $k$ can be found recursively, and
parametrized in the same way as the wavefunctions, eq.\
(\ref{wfn}), using the partitions of $N$ and $n$. First,
the total number of states can be written as %
\be%
\#_{N,N_\phi,k}= \sum_{{m_1,\ldots,m_k:} {\sum_\alpha \alpha
m_\alpha=N}}\prod_\alpha{m_\alpha+n_\alpha \choose n_\alpha}.\ee
Here the recursive structure is clear; notice that $m_k=(N-F)/k$,
$n_{k-1}=n'$, $m_{k-1}=(F-F')/k'$, etc., so this agrees with
preceding results. In practice we need the values of $n_\alpha$
for given $N$ and $n$; these are obtained by solving the
conditions on the number of $w^{(\alpha)}_l$ at each level
$\alpha$ in terms of the
$m_\beta$'s:%
\be%
n_\alpha=\alpha\left(\frac{2N+n}{k}-\sum_{\beta:\beta\geq\alpha}
2m_\beta\right)
-\sum_{\beta<\alpha}2\beta m_\beta;\ee%
for $\alpha=k$, this reduces to $n_k=n$. There is a similar form
for the coefficients
$\left\{\begin{array}{c}n\\F\end{array}\right\}_k$, as a sum over
partitions of $F=N-km_k$, which agrees with the result of Ref.\
\cite{ardonne}.

One further refinement is to count zero-energy eigenstates for
each angular momentum value $L_z$ on the sphere. Instead of ${\rm
dim}\, V_{N,N_\phi,k}$, we can choose to calculate the trace of
$q^{NN_\phi/2-L_z}$, where $q$ is an indeterminate, and $L_z$ is
the $z$-component of total angular momentum $\bf L$. We recall
that for the single-particle wavefunctions, the function $z^m$
contributes $N_\phi/2-m$ to $L_z$, for $m=0$, \ldots, $N_\phi$
\cite{hald83,rr96}. Thus (for $L_z$ eigenstates) we have arranged
that the exponent of $q$ is simply the total degree of the
wavefunction in the $z$'s. {}From the number of states of each
$L_z$, one can extract the number of quasihole states of each
value of ${\bf L}^2$ also. This information can be valuable in
numerical studies.

Once again it is sufficient to count residues, this time of each
degree. The total degree in the cluster coordinates $Z_p$, $p=1$,
\ldots, $m_k$, is $km_k(m_k-1)$ plus the total degree of the
symmetric polynomial in the $Z$'s produced by expanding the factor
$\prod_{p,l}(Z_p-w_l)$, plus the contributions of the factors that
connect with the unclustered particles. The number of symmetric
polynomials of degree $d$ in $m_k$ variables and of degree at most
$n_k$ in each variable is given by the coefficient of the $q^d$
term in
the (Gauss) $q$-binomial coefficient (a polynomial in $q$), defined by%
\be%
{m_k+n_k\choose
n_k}_{\!q}=\frac{(m_k+n_k)_q!}{(m_k)_q!(n_k)_q!},\ee%
where the $q$-factorial $(n)_q!=(n)_q(n-1)_q\cdots(1)_q$, and the
$q$-deformed integers are%
\be%
(n)_q=1+q+q^2+\cdots+q^{n-1}=\frac{1-q^n}{1-q}.\ee%
This combinatorial result can be understood from its own
generating function, which is%
\be%
\frac{1}{(1-x)(1-xq)\cdots(1-xq^{m_k})}=\sum_{n_k=0}^\infty
{m_k+n_k\choose n_k}_{\!q} x^{n_k};\label{genfn}\ee%
this identity is Heine's $q$-binomial formula \cite{kc}. It is
easily seen that the left-hand side of eq.\ (\ref{genfn}) is the
generating function for symmetric polynomials in $m_k$ variables,
of degree at most $n_k$ in each variable: The symmetric
polynomials in $m_k$ variables $Z_p$ are (freely) generated by the
elementary symmetric polynomials $e_1=\sum_p Z_p$,
$e_2=\sum_{p<p'}Z_pZ_{p'}$, \ldots, $e_{m_k}=\prod_p Z_p$, which
are of total degrees $1$, \ldots, $m_k$, respectively, and of
degree at most 1 in each variable. Each factor $(1-xq^m)^{-1}$
gives a series that contains one term $x^rq^{mr}$ with coefficient
$1$, corresponding to each distinct possible factor $e_m^r$.
Similar formulas apply to each level in the recursion, with
$\alpha$ in place of $k$. In all these expressions, the limit
$q\to1$ reproduces earlier or well-known results, with ordinary
binomials, factorials, etc., in place of their $q$-analogues.

Since the $q$-binomials count the effects of the factors
containing $w^{(\alpha)}_l$'s, we now only require the total
degree (summed over $\alpha$) of the minimal degree polynomial in
the $Z$'s of all levels, that is the term of highest possible
degree in all $w$'s. For the given partition described by
$m_\alpha$, this minimal
total degree is%
\bea%
\lefteqn{\sum_\alpha \alpha m_\alpha(m_\alpha-1)+
\sum_{\alpha<\beta}2\alpha m_\alpha m_\beta,}\qquad\qquad&&\nonumber\\
&=&
\sum_{\alpha,\beta}m_\alpha m_\beta M_{\alpha\beta}-N,\eea %
where $M_{\alpha\beta}={\rm min}\,(\alpha,\beta)$ \cite{sf,aks}.

The result is that the trace over $V_{N,N_\phi,k}$ is%
\widetext 
\be%
{\rm tr}\,q^{NN_\phi/2-L_z}=q^{-N}\sum_{{m_1,\ldots,m_k:}
{\sum_\alpha \alpha m_\alpha=N}}q^{\sum_{\beta,\gamma}m_\beta
m_\gamma M_{\beta\gamma}} \prod_\alpha{m_\alpha+n_\alpha \choose
n_\alpha}_{\!q}.\label{fincount}\ee %
This expression counts the zero-energy states on the finite-size
sphere for each $L_z$; furthermore, ${\rm tr}\, q^{-L_z}$ for
$M\geq0$ is independent of $M$. As a test, this formula has been
compared with some rows for $k>2$ in the tables of numbers of
states for each $L$ that have appeared in Ref.\ \cite{gr} (for
$k=3$) and recently in Ref.\ \cite{rj}, with perfect agreement
\cite{nigel}. The corresponding result that generalizes
$\left\{\begin{array}{c}n\\F\end{array}\right\}_k$ for $q\neq1$
was given in Ref.\cite{ardonne} as the ``truncated character'' for
parafermions (but of course does not describe the total orbital
angular momentum of the states). Those expressions now emerge from
analyzing polynomials in $z$'s.
\bottom{-2.7cm} \narrowtext %

It is possible to take the limit $N_\phi\to\infty$ with $m_\alpha$
fixed. In this limit, all $n_\alpha\to \infty$,
and by assuming that $|q|<1$,%
\be%
{m_\alpha+n_\alpha \choose
n_\alpha}_{\!q}\to\prod_{m=1}^{m_\alpha}\frac{1}{(1-q^{m})}.\ee %
Then the resulting generating function for the number of
zero-energy wavefunctions with no upper limit on the degree in
each variable is identical with the results of Refs.\
\cite{sf,aks}. These references also explain how to shift the
degree in $q$ so that the subsequent limit $N\to\infty$ can be
taken for each term of fixed degree in $q$. For the simplest case,
one divides by $q^{N^2/k-N}$, and the limit counts the number of
excitations at finite change in angular momentum of the edge of a
very large drop of particles, similar to the results of Ref.\
\cite{milr} for the paired states (the $k=2$ case). In this limit
the leading term in the generating function when $N$ is divisible
by $k$ is $q^0$ and corresponds to the vacuum of the conformal
field theory for the edge, as there are no quasiholes in the
interior. More generally, one may also take the $N_\phi$ and
$N\to\infty$ limits with some quasiholes held fixed at the center
of the drop, and arrange that the powers of $q$ in general are the
conformal weights of the states on the edge. For up to $k$
quasiholes at the center, this was done in Refs.\ \cite{sf,aks},
and produces the character of the chiral conformal field theory
for each sector of edge states (there are $k+1$ distinct sectors
for the $M=0$ case treated here, but this number does depend on
$M$ \cite{rr99}).

To conclude, we now have a satisfactory understanding of the trial
quasihole wavefunctions of the RR states. We expect that the
approach can be generalized to obtain analogous results for the
quasiholes of other quantum Hall phases that have trial states
that are zero-energy eigenstates of short-range interactions, such
as bilayer states, spin-polarized ground states of spin-1/2
particles, and the various spin-singlet clustered states for
particles with spin 1/2 or 1 that have been introduced in the
literature; the results of Ref.\ \cite{aks2} are relevant to the
latter. We want to emphasize that ways of writing the quasihole
wavefunctions other than the one presented here could still be of
potential interest for the purpose of gaining useful physical
insight, or for performing further calculations of the properties
of the quasiholes.


{\it Note added}: After submission of the original version of this
paper, some additional facts have come to light which it may be
useful to describe here for completeness. First, I have become
aware that an earlier, difficult paper by Feigin and Loktev
obtains some of the results herein, including the counting
formula, eq.\ (\ref{fincount}). In addition, there is a connection
with other recent work on the zero-energy states for the
$k+1$-body Hamiltonian. Several authors \cite{haldmm,bkwhk,sl}
(some of whom concentrated on $k=2$ only) have discovered that for
the system on a torus or cylinder, as the radius is taken to zero,
an effective description emerges that is combinatorially simple.
[This builds on earlier corresponding results for the Laughlin
Abelian ($k=1$) states, and others \cite{rhsl}.] For the cylinder
(which topologically is the same as the sphere), we will consider
the single-particle orbitals (corresponding to single-particle
$L_z$ eigenstates $z^m$ on the sphere) to be labeled by $m+1=1$,
\ldots, $N_\phi+1$, arranged in sequence on a line. Then the
recipe is that, for $M=0$, the Hamiltonian stipulates that in a
zero-energy state, any two neighboring orbitals can be occupied by
a total of not more than $k$ particles, which can be distributed
in any way between the two orbitals. (For the $M=1$ case, the
first one relevant to fermions, this becomes no more than $k$ in
any group of $k+2$ neighboring orbitals, with of course not more
than one in each orbital because of the Pauli exclusion
principle.) For the torus, the same Hamiltonian applies for
orbitals with periodic boundary conditions, $N_\phi+1\equiv 1$,
with ``neighboring orbitals'' interpreted cyclically. Clearly,
either of these specifies a combinatorial problem in which one
would like to count the total number of such states for each $N$,
$N_\phi$, $k$, and one would hope that the result for the sphere
(cylinder) agrees with the formula, eq.\ (\ref{fincount}),
obtained here.

This type of problem is connected with a long history of results
in combinatorics, for which see Andrews' book \cite{andrews}. The
problem for bosons in the cylinder/sphere case can be related to
counting partitions, by identifying the length of each row of (the
Ferrers-Young diagram of) a partition with the orbital number
$m+1=1$, $2$, \ldots, of a corresponding particle. As partitions
are defined to have rows of (not strictly) decreasing lengths,
each partition corresponds to a unique many-particle basis state
for a system of bosons, the number $N$ of bosons being the number
of rows of non-zero length in the partition. (Similarly, for
fermions there is a correspondence with partitions with all rows
having distinct lengths. There is a correspondence between the two
problems, obtained by adding $N-i$ to the length of the $i$th row
of a partition for the boson case to obtain the corresponding
partition for fermions.) Notice that the total number of boxes
${\cal N}$ (the number being partitioned) is related to the total
$L_z$ by $NN_\phi/2+N-L_z={\cal N}$ for $M=0$. Then in the limit
$N_\phi\to\infty$, in which there is no upper limit on lengths of
rows, the simplest case is the $k=1$ case for bosons. The
zero-energy states then correspond to partitions in which two
neighboring rows have lengths differing by at least two. The
number of these is described by a generating function that can be
written in two ways; this is the celebrated Rogers-Ramanujan
identity, which is thus connected combinatorially with the
quasiholes of the Laughlin $\nu=1/2$ state in the plane. The
generalization to $k>1$ is given by the Gordon-Andrews identities,
see Ref.\ \cite{andrews}. One side of each of these identities
agrees with the $N_\phi\to\infty$ limit of our result, studied
earlier in \cite{sf,aks}. (There are further results also, in
which some number $l$, $0\leq l\leq k$ of quasiholes are fixed at
the origin \cite{sf,aks}. In the chain of orbitals, this
corresponds to a restriction that the occupation of the first
orbital is $\leq k-l$.)

To make full contact with our results, we require results for
partitions with the additional restriction of a maximum length for
the rows. For the $k=1$ case, this ``finitization'' of the
Rogers-Ramanujan identities was found in Ref.\ \cite{andrews70}.
For $k>1$, the number of partitions obeying the given rules, and
the appropriate finitization of the Gordon-Andrews identities, was
derived in detail by Warnaar \cite{warnaar} in a manner very close
to the above interpretation as particles in orbitals. [The
identities are also given in Ref.\ \cite{fl}, in a ``grand
canonical'' form in which one multiplies the results for fixed $N$
by $x^N$ (where $x$ is another indeterminate) and then sums over
$N$ with $N_\phi$ fixed.] Here boundary conditions can be imposed
at the two ends of the chain, that the occupation of the first
orbital is $\leq k-l$, and that of the last is $\leq k-l'$. For
the quantum Hall states on the sphere, these correspond to fixing
$l$ quasiholes at the north pole, $l'$ at the south pole. One side
of each of these identities (specialized to the case $l=l'=0$)
agrees with the formula (\ref{fincount}) (times $q^N$). These
results thus confirm that the counting produced by these simple
one-dimensional Hamiltonians \cite{rhsl,haldmm,bkwhk,sl}
reproduces that obtained from the two-dimensional analysis here.

For the torus, the periodic boundary condition on the
one-dimensional system can also be handled. The ends of the open
chain can be joined and the restriction of total occupation $\leq
k$ for the pair of orbitals $1$, $N_\phi$ can be imposed by taking
Warnaar's formulas, setting $l+l'=k$, and summing over $l$ to
obtain the total number of zero-energy states. It is an open
problem to derive this using the wavefunctions on the torus in a
fashion similar to that of the present paper.

\vspace{0.2in}

I am grateful to N.R. Cooper for many stimulating discussions and
for pointing out several errors in the manuscript, and to R. Howe
and I.B. Frenkel for helpful remarks. This work was supported by
NSF grant no.\ DMR-02-42949.

\vspace{0.2in}


{\bf Appendix A}: The proof that a symmetric polynomial that is
non-zero when $z_1=z_2=\cdots =z_k$, and zero when in addition
$z_{k+1}=z_1$, vanishes quadratically $(z_{k+1}-z_1)^2$ (or
faster) as $z_{k+1}\to z_1$, goes as follows. We will work on the
sphere \cite{hald83}, using the homogeneous or spinor particle
coordinates $u_i$, $v_i$, in terms of which $z_i=v_i/u_i$ for each
$i$. We will use auxiliary coordinates like the $w_l$ in eq.\
(\ref{res}) to form coherent states; these correspond to spinors
$\alpha_l$, $\beta_l$ with $w_l=\beta_l/\alpha_l$ (examples of
such functions appear in the main text). In this form, any
wavefunction is rotationally invariant if the rotation acts on all
of the $\alpha_l$, $\beta_l$ pairs as well as on all the $u_i$,
$v_i$ pairs. That is, the wavefunction is a sum of products of
factors of the forms $u_iv_j-v_iu_j$ (or similarly with
$\alpha_l$, $\beta_l$ pairs in place of one or both particle
spinors). This is because this singlet combination of two spinors
is the only way to construct invariant functions out of the $N$
spinors $u_i$, $v_i$ and auxiliary spinors $\alpha_l$, $\beta_l$.
For our purposes, the polynomials must be homogeneous of degree
$N_\phi$ in each $u_i$, $v_i$ (corresponding to each particle
having angular momentum $N_\phi/2$), and also symmetric under
permutations of the particles. Then a wavefunction can be analyzed
into terms according to the angular momentum $\bf L$ of some set
of $k+1$ particles, schematically
$\Psi=\sum_{L,L_z}\Psi_{L,L_z}(k+1)\Psi_{L,-L_z}(N-k-1)$, where
$\Psi_{L,L_z}(k+1)$ is a function of the first $k+1$ coordinates
only, and $\Psi_{L,-L_z}(N-k-1)$ is a function of the remaining
particle and all the auxiliary coordinates. These are combined so
that the state $\Psi$ is a singlet. If the state
$\Psi_{L,L_z}(k+1)$ for $L=L_{\rm max}=(k+1)N_\phi/2$ is nonzero,
then in particular its component with $L_z=L$ is nonzero, and this
is represented by $\prod_i u_i^{N_\phi}$, so that it is nonzero
when all $v_i=0$ (i.e.\ $z_1=\cdots=z_{k+1}=0$). Hence we require
these $L=L_{\rm max}$ components to vanish. The rate at which the
function vanishes as $z_{k+1}\to z_1=z_2=\cdots =z_k=0$ is
determined by the largest angular momentum value $L$ for which the
component $\Psi_{L,L_z}(k+1)$ is nonzero. By analyzing the totally
symmetric states of $k+1$ spins of magnitude $N_\phi/2$, one finds
that the largest values are $L_{\rm max}$, $L_{\rm max}-2$, \ldots
(the multiplicities are one for these first two terms, but larger
than one for subsequent terms). Thus the first subleading term
contains factors $(u_iv_j-v_iu_j)^2$, or $(z_i-z_j)^2$, which is
what we wanted to prove. This proof uses the sphere, but can
presumably be repeated directly for the plane by reference to the
center of mass as well as the total degree. However, this is
unnecessary, as the counting of states for the plane subject to an
upper limit $N_\phi$ on the degree in each variable is always the
same as for the sphere.

{}From this result, it follows that if coordinates $z_{k+1}$,
$z_{k+2}$, \ldots, $z_{2k}$ approach $z_1=\cdots=z_k$, then the
wavefunction must vanish quadratically in each of the second $k$
coordinates, and so if $z_{k+1}=\cdots =z_{2k}$, it vanishes as
$(z_1-z_{k+1})^{2k}$ as $z_{k+1}\to z_1$, as was also used in the
form of the residues.

\vspace{0.2in}

{\bf Appendix B}: Here we describe some alternative quasihole
wavefunctions, which are natural generalizations of the ground
state and quasihole wavefunctions in the form given by Cappelli
{\it et al.} \cite{cgt} (rather than those of Ref.\ \cite{rr99}).
In these partitions of $N$ and $n$ are again used, but are more
conveniently labeled in a different, but related way: $N$ (resp.,
$n$) is partitioned into $N_1\geq N_2\geq\cdots\geq N_k$ (resp.,
$\hat{n}_1\leq \hat{n}_2\leq\cdots\leq \hat{n}_k$) with
$\sum_{\alpha=1}^kN_\alpha=N$ (resp., $\sum_{\alpha=1}^k
\hat{n}_\alpha=n$). The relation to the numbers used in the main
text is $m_\alpha=N_\alpha-N_{\alpha+1}$ ($N_{k+1}=0$),
$n_\alpha=\sum_{\beta:\beta\leq\alpha}\hat{n}_\beta$. The
wavefunctions are functions of the $z$'s and only one set $w_1$,
\ldots, $w_n$ of quasihole coordinates. The
coherent state wavefunctions are given by:%
\widetext \top{-2.8cm}
\be%
\widetilde{\Psi}_{N,N_\phi,k}(z_1,\ldots,z_N;w_1,\ldots,w_n)={\cal
S}_z\left\{{\prod_{i<j}}^{(1)}(z_i-z_j)^2\cdots{\prod_{i<j}}^{(k)}(z_i-z_j)^2
{\prod_{i,l}}^{(1)}(z_i-w_l)\cdots{\prod_{i,l}}^{(k)}(z_i-w_l)
\right\}.\label{wfnc}\ee%
\bottom{-2.7cm} \narrowtext \noindent%
In the products $\prod^{(\alpha)}$ (for $\alpha=1$, \ldots, $k$),
$i$, $j$ run over $\sum_{\beta:\beta<\alpha}N_\beta+1$, \ldots,
$\sum_{\beta:\beta\leq\alpha}N_\beta$, and also $l$ runs over
$\sum_{\beta:\beta<\alpha}\hat{n}_\beta+1$, \ldots,
$\sum_{\beta:\beta\leq\alpha}\hat{n}_\beta$. To ensure that each
function is of the same degree in all coordinates $z_i$, we impose
$2N_\alpha+\hat{n}_\alpha=(2N+n)/k$ for all $\alpha$ (which are
equivalent to the earlier conditions). Note that each function
need not be of the same degree in all the $w$'s, and one should
not symmetrize over the $w$'s. It is easy to see that when any
$k+1$ $z$'s are equal, the function vanishes, because in any term
in the symmetrization over the $z$'s, at least two must be of the
same ``type'' $\alpha$. In the limit as $z_{(p-1)k+1}$, \ldots,
$z_{(p-1)k+k}\to Z_p$ (for $p=1$, \ldots, $N_k=m_k$), the function
vanishes if a further set of $k$ coordinates coincide with one
another. Thus each of them reduces to the form of a non-vanishing
residue with $F=N-kN_k$, but in which
$\widetilde{\Psi}_{F,N_\phi',k'}$ is of the form (\ref{wfnc}) with
the parameters as given, and which, unlike the residues above,
still depends on the first $n-\hat{n}_k$ $w_l$'s as well as $F$ of
the $z$'s. In this function $\widetilde{\Psi}_{F,N_\phi',k'}$, the
partition of $N$ is replaced by a similar partition of $F$, with
$F_\alpha=N_\alpha-N_k$ ($\alpha=1$, \ldots, $k'$), while the
partition $\hat{n}_\alpha$ of $n$ is unchanged, except that
$\hat{n}_k$ must be dropped.

The fact that $\widetilde{\Psi}_{F,N_\phi',k'}$ here depends on
some of the same $w$'s as appear in front of it in the residue
makes it nontrivial to verify that each nonvanishing residue can
be obtained from some (linear combination of) functions of this
form. One would like to expand the non-vanishing residues that
result from eq.\ (\ref{wfnc}) in linearly-independent polynomials
in $w$'s, in order to find the dimension of the space spanned by
the coherent states. The clustered part of the residue produces a
linearly-independent set of ${(N-F)/k+n \choose n}$ symmetric
polynomials in the $w$'s, while the remaining factor
$\widetilde{\Psi}_{F,N_\phi',k'}$ would be assumed by induction to
be spanned by some linearly-independent set of $\#_{F,N_\phi',k'}$
polynomials in the $w$'s. However, we were not able to establish
that the total number of independent polynomials in $w$'s in the
products of these two sets is given by eq.\ (\ref{rec}). The
reason is that the set of polynomials in $w$'s obtained by
multiplying one member of one of these two sets of polynomials by
a member of the other may not be linearly independent, even though
both the original sets were. (This may be formulated
mathematically as asking whether the set that spans the expansion
of $\widetilde{\Psi}_{F,N_\phi',k'}$ is linearly-independent {\em
over the algebra of symmetric polynomials}, not just over the
complex numbers.) We were not able to establish the linear
independence of the resulting set in general, or that the number
of independent functions equals the required number, though in
special cases of small systems it does \cite{nigel}. If
$\widetilde{\Psi}_{F,N_\phi',k'}$ could be expanded in ${\rm
S}_n$-harmonic polynomials in $w$'s, then the argument would go
through. As usual, ${\rm S}_n$ denotes the symmetric (or
permutation) group on $n$ objects. The ${\rm S}_n$-harmonic
polynomials are a linearly-independent set of $n!$ polynomials in
the $n$ variables $w_l$ which, when multiplied by arbitrary
symmetric polynomials, span the full space of polynomials in $w$'s
(thus, they form a basis for the latter vector space over the
algebra of symmetric polynomials) \cite{roger}. Though, in the
cases we checked, there are enough ${\rm S}_n$-harmonic
polynomials in each degree for this to work (and one must note
that permutations of the $w_l$'s in the wavefunctions (\ref{wfnc})
leave the functions of $z$'s invariant, so only one member of each
orbit under ${\rm S}_n$ in the ${\rm S}_n$-harmonic polynomials
can be used), we do not know if the functions can always be
expanded in this way. Thus, while we suspect that the quasihole
wavefunctions in eq.\ (\ref{wfnc}) do span the space
$V_{N,N_\phi,k}$ for all $N$, $N_\phi$, $k$, we have been unable
to prove this so far.


\widetext


\vspace{-0.25in}

\begin{references}

\vspace{-0.5in}

\bibitem{mr}
G.~Moore and N.~Read, Nucl.\ Phys.\ {\bf B 360}, 362 (1991).

\bibitem{rr99}
N. Read and E. Rezayi, \prb {\bf 59}, 8084 (1999).

\bibitem{cgt} A. Cappelli, L.S. Georgiev, and I.T. Todorov, Nucl.
Phys. B {\bf 599}, 499 (2001).

\bibitem{nayak} C.~Nayak and F.~Wilczek, Nucl. Phys. B{\bf 479}, 529 (1996).

\bibitem{rr96}
N. Read and E. Rezayi, Phys. Rev. B {\bf 54}, 16864 (1996).

\bibitem{gr} V. Gurarie and E. Rezayi, Phys. Rev. B {\bf 61}, 5473
(2000).

\bibitem{ardonne} E. Ardonne, J. Phys. A {\bf 35}, 447 (2002).

\bibitem{rj} N. Regnault and T. Jolicoeur, cond-mat/0601550.

\bibitem{morfrh} R.H. Morf, Phys. Rev. Lett. {\bf 80}, 1505 (1998);
E.H. Rezayi and F.D.M. Haldane, Phys. Rev. Lett. {\bf 84}, 4685
(2000).

\bibitem{cwg}
N.R. Cooper, N.K. Wilkin, and J.M.F. Gunn, Phys.\ Rev.\ Lett.\
{\bf 87}, 120405 (2001).

\bibitem{rrc} E.H. Rezayi, N. Read, and N.R. Cooper, Phys. Rev. Lett.
{\bf 95}, 160404 (2005).

\bibitem{freedman} M.H. Freedman, A. Kitaev, M.J. Larsen, and
Z. Wang, quant-ph/0101025.

\bibitem{hald83}
   F.D.M. Haldane, Phys.\ Rev.\ Lett.\ {\bf 51}, 605 (1983).

\bibitem{sf} A.V. Stoyanovsky and B.L. Feigin, Funct. Anal. Appl.
{\bf 28}, 55 (1994); B. Feigin, M. Jimbo, R. Kedem, S. Loktev, and
T. Miwa, J. Alg. {\bf 279}, 147 (2004).

\bibitem{aks} E. Ardonne, R. Kedem, and M. Stone, J. Phys. A {\bf 38},
617 (2005).

\bibitem{arrs} E. Ardonne, N. Read, E. Rezayi, and K. Schoutens,
Nucl. Phys. B {\bf 607}, 549 (2001).

\bibitem{schout} K. Schoutens, Phys. Rev. Lett. {\bf 79}, 2608
(1997).

\bibitem{kc} This formula can be proved by induction on $m_k$,
or by other methods; see e.g.\ V. Kac and P. Cheung, {\it Quantum
Calculus} (Springer, New York, NY, 2002).

\bibitem{nigel} N.R. Cooper, private communication.

\bibitem{milr} M. Milovanovic and N. Read, Phys. Rev. B {\bf 53}, 13559
   (1996).

\bibitem{aks2} E. Ardonne, R. Kedem, and M. Stone,
J. Phys. A {\bf 38}, 9183 (2005); math.RT/0504364.

\bibitem{fl} B.L. Feigin and S.A. Loktev, Funct. Anal. Appl. {\bf
35}, 44 (2001).

\bibitem{haldmm} F.D.M. Haldane, talk at APS March Meeting,
Baltimore, March, 2006.

\bibitem{bkwhk} E.J. Bergholtz, J. Kailasvuori, E. Wikberg, T.H.
Hansson, and A. Karlhede, cond-mat/0604251.

\bibitem{sl} A. Seidel and D.-H. Lee, cond-mat/0604465.

\bibitem{rhsl} E.H. Rezayi and F.D.M. Haldane, Phys. Rev. B. {\bf
50}, 17199 (1994); E.J. Bergholtz and A. Karlhede, Phys. Rev.
Lett. {\bf 94}, 26802 (2005); cond-mat/0509434; A. Seidel, H. Fu,
D.-H. Lee, J.M Leinaas, J. Moore, Phys. Rev. Lett. {\bf 95},
266405 (2005).

\bibitem{andrews} G.E. Andrews, {\it The Theory of Partitions},
(Cambridge University Press, Cambridge, 1984), especially Ch.\ 7.

\bibitem{andrews70} G.E. Andrews, Scripta Math. {\bf 28}, 297
(1970), or see pp. 50, 157 in Ref.\ \cite{andrews}.

\bibitem{warnaar} S.O. Warnaar, Commun. Math. Phys. {\bf 184}, 203
(1997).

\bibitem{roger} R. Howe, private communication.

\end{references}
\end{document}